# Fubini–Study Speed and Short-Time Identification of Bosonic Instability

A. M. Tishin[a,b]*

[a] Lomonosov Moscow State University, Physics Department, 119991, Leninskie gory 1, Moscow, Russia

[b] Moscow Institute of Physics and Technology, 141701, Institutskiy per. 9, Dolgoprudny, Moscow Region, Russia

*tishin@amtc.org

## Abstract

We ask which features of dynamical instability in quadratic bosonic systems can be inferred from short-time quantum distinguishability when the response depends on the input state. For a fixed autonomous quadratic Hamiltonian, we show that the infimum of the Fubini–Study speed over centered pure Gaussian inputs equals the Euclidean norm of the positive real parts of the flow eigenvalues divided by $\sqrt{2}$. The infimum is attained exactly when the flow generator is diagonalizable. For each input, the excess squared speed equals one eighth of the squared Henrici departure from normality in the adapted quadratures. For a detuned parametric amplifier, we derive the exact minimum speed under a finite squeezing budget. In the lossy single-mode amplifier, the initial slope and curvature of the bare-vacuum return fidelity are independent of detuning. With a common phase-insensitive Markovian bath, adding a return trace from a squeezed vacuum with calibrated squeezing strength and angle determines the growth margin, whose sign identifies the side of the instability threshold. The same data determine the damping and diffusion coefficients. Finite-window simulations illustrate discrimination between representative cases on opposite sides of the threshold.

.

## 1. Introduction

The Fubini–Study metric quantifies the infinitesimal distinguishability of pure quantum states [1]. For unitary evolution the speed is the energy uncertainty in frequency units [2]. Its mixed-state analogue is set by the quantum Fisher information [3–4]. The speed also enters the Mandelstam–Tamm bound [5] and related quantum speed limits [6].

A quadratic bosonic Hamiltonian supplies a second rate, the largest real part of the symplectic flow spectrum. The two rates coincide up to a factor for a resonant squeezer. Links between quantum speed limits and classical instability have also been found in particular Gaussian state-preparation protocols [7]. The energy uncertainty depends on the input state, however, whereas the flow spectrum is fixed by the Hamiltonian. This difference makes state preparation a control on the geometric speed at a fixed instability spectrum.

Gaussian quantum Fisher information separates changes of the symplectic spectrum from correlation-generating dynamics [8]. The bare vacuum is invariant under passive Gaussian unitaries, a property specific to that reference. Independently, classical stability estimates bound instantaneous growth through the symmetric part of the generator [9]. Other connections between quantum information and instability concern entropy or coherence growth in Gaussian environments [10] and speed limits for out-of-time-ordered correlators [11]. Non-normality has been studied in closed quadratic bosonic systems [12] and quadratic Markovian extensions [13–16]. Equal Lyapunov exponents can accompany different finite-time gain in collective-spin models [17]. In photonic time crystals the Petermann factor governs the photon-number prefactor while the Floquet exponent sets the growth timescale [18]. Symplectic normal forms classify stable, unstable and defective quadratic generators [19]. For the intrinsic geometry of centered pure Gaussian states [20], the spectral form of our infimum has a classical analogue in the translation length of symplectic maps on Siegel space [21].

Here we determine the exact speed floor at fixed Hamiltonian. An identity for the entire unstable spectrum separates the invariant contribution from a state-dependent departure from normality, and the infimum fixes when preparation can remove that excess. The amplifier then gives two complementary results. Finite input squeezing imposes an exact constraint on the speed near threshold. A rotated squeezed input supplies the missing detuning information in a short-time return protocol and separates it from isotropic damping and diffusion.

## 2. Setup

For pure states the quantum geometric tensor [1] gives $v_{\mathrm{FS}} = ds_{\mathrm{FS}}/dt$. We use squared fidelity, so $1 - F(t, t + dt) = v_{\mathrm{FS}}^2\, dt^2 + o(dt^2)$. In unitary evolution $v_{\mathrm{FS}} = \Delta\hat{\mathrm{H}}/\hbar$ [2]. The mixed-state Bures speed is $\tfrac{1}{2}\sqrt{F_{\mathrm{Q}}}$ [3–4]. For unitary encoding by $\hat{\mathrm{H}}$, $F_{\mathrm{Q}} \leq 4(\Delta\hat{\mathrm{H}})^2/\hbar^2$. Closed Gaussian expressions are available [22–24].

For $N$ bosonic modes write $\hat{z} = (\hat{x}_1, \hat{p}_1, \ldots, \hat{x}_N, \hat{p}_N)$, the canonical symplectic form $J$ with $J^2 = -1$, and

$$\hat{\mathrm{H}} = \tfrac{1}{2}\hat{\mathrm{z}}^{\mathrm{T}} G\hat{\mathrm{z}}, \quad M = JG \in \mathrm{sp}(2N, \mathbb{R}) \tag{1}$$

with $G$ real symmetric, in frequency units ($\hbar = 1$). The Heisenberg equations are $d\hat{z}/dt = M\hat{z}$. Eigenvalues of $M$ come in ± pairs; $\lambda_{\max}$ denotes the largest real part. The bare-mode vacuum, the vacuum of the quadratures in which $G$ is written, has covariance $\Sigma = \tfrac{1}{2}\cdot 1$ [25–26].

A passive change of coordinates $\hat{\mathrm{z}} \to V^{-1}\hat{\mathrm{z}}$ with symplectic $V$ transforms both the generator and the covariance, with $G \to V^{\mathrm{T}} G V$ and $M \to V^{-1} M V$. It leaves the physical speed unchanged. An active change of generator at fixed bare vacuum instead compares different Hamiltonians with the same spectrum. Equivalently, one can hold the Hamiltonian fixed and change the prepared covariance to $VV^{\mathrm{T}}/2$. We use the latter interpretation for the optimization in Sec. 5.

The bare-mode vacuum is specified independently of the stability of $M$. It has finite energy variance for every finite real symmetric $G$, including in the hyperbolic regime where an instantaneous normalizable eigenvacuum can fail to exist [27]. The speed here is evaluated along the trajectory of a chosen state. The symplectic quantum geometric tensor of Bogoliubov modes is defined on the mode space [28].

**3. Decomposition and sharp instability bound**

*Theorem 1.* For every $N$ and every real symmetric $G$, the Fubini–Study speed at the bare-mode vacuum obeys

$$8v_{\mathrm{FS}}^2 = \mathrm{Tr}(G^2) + \mathrm{Tr}(M^2) = 2\|S\|_{\mathrm{F}}^2, \quad S = (M + M^{\mathrm{T}})/2 \tag{2}$$

with $\|\cdot\|_{\mathrm{F}}$ the Frobenius norm, so $v_{\mathrm{FS}} = \frac{1}{4}\|M + M^{\mathrm{T}}\|_{\mathrm{F}}$ depends on the symmetric part of the flow alone.

*Proof.* For a centered Gaussian state of covariance $\Sigma$, $\mathrm{Var}(\hat{H}) = \frac{1}{2}\,\mathrm{Tr}(G\Sigma G\Sigma) + \frac{1}{8}\,\mathrm{Tr}(GJGJ)$. At $\Sigma = \frac{1}{2}\cdot 1$ the first term is $\frac{1}{8}\,\mathrm{Tr}\, G^2$; since $M^2 = JGJG$ and the trace is cyclic, the second is $\frac{1}{8}\,\mathrm{Tr}\, M^2$. The reference state is pure, so $v_{\mathrm{FS}} = \sqrt{\mathrm{Var}(\hat{H})}$ [2]. For the second equality, $M^{\mathrm{T}}M = G^2$ gives $\mathrm{Tr}\, G^2 = \|M\|_{\mathrm{F}}^2$; splitting $M = S + A$ into symmetric and antisymmetric parts and using $\mathrm{Tr}(SA) = 0$ gives $\|M\|_{\mathrm{F}}^2 = \|S\|_{\mathrm{F}}^2 + \|A\|_{\mathrm{F}}^2$ and $\mathrm{Tr}\, M^2 = \|S\|_{\mathrm{F}}^2 - \|A\|_{\mathrm{F}}^2$, so the antisymmetric contributions cancel. ■

Neither equality is new algebra. What they supply is a reading that the first form conceals: the antisymmetric part of the flow rotates, the symmetric part stretches, and at the vacuum the geometric speed measures the second and is completely blind to the first, whereas the eigenvalues of $M$ depend on both because rotation competes with stretching in setting the growth rate.

The blindness is a statement about which Hamiltonians are invisible. The passive, photon-number-conserving sector is characterized by a symmetric $G_{\mathrm{p}}$ with $[J, G_{\mathrm{p}}] = 0$ — detunings, phase shifters, beam splitters. Its contribution to the flow is purely antisymmetric, since $M_{\mathrm{p}} + M_{\mathrm{p}}^{\mathrm{T}} = [J, G_{\mathrm{p}}] = 0$, so by Eq. (2) it contributes nothing to $v_{\mathrm{FS}}$ while entering the spectrum of $M$. That a purely passive generator produces no geometric motion of the vacuum reference used here follows from the even–odd decomposition of the Gaussian quantum Fisher information [8] and is not claimed here as new; Eq. (2) is stronger only in the additive sense used below, that a passive term added to an active generator produces no cross term in $v_{\mathrm{FS}}$ either. Hence:

*Corollary 1.* Generator–state pairs with identical flow spectra can have parametrically different $v_{\mathrm{FS}}$; the explicit isospectral family constructed in the Supplemental Material is unbounded within the quadratic model and already changes $v_{\mathrm{FS}}$ by a factor 11.1 over the range shown, with every exponent unchanged by construction, since the family acts by similarity on $M$, and with numerical variation below $2.2 \times 10^{-16}$ [29].

The construction is the following. At $N = 2$ the bare-vacuum covariance is held fixed at $\Sigma = \frac{1}{2}\cdot 1$ while the generator is replaced along a one-parameter family of active symplectic squeezes, active in the sense fixed above. The resulting flow matrices are similar to one another and therefore share an identical spectrum, so the family consists of physically distinct generator–state pairs and not of one system written in different coordinates. Moving the inverse transformation from the generator onto the state, the same algebra reads equivalently as one fixed generator probed on a family of squeezed Gaussian references; the convention used here holds the reference fixed and transforms the generator. Random sampling corroborates the conclusion without relying on the construction at all: among two-mode generators whose geometric speed agrees to within 2 %, the largest exponent covers essentially the whole interval permitted by Theorem 2, from numerical zero up to 0.99 of the ceiling [29].

The factor 11.1 is specific to the plotted range. In the explicit family the Frobenius norm of the generator grows without bound along the symplectic similarity while its spectrum remains fixed. Equation (2) then makes the speed unbounded as well. This comparison varies the physical Hamiltonian at fixed bare vacuum, or equivalently varies the input state at fixed Hamiltonian.

*Corollary 2.* At the fixed bare-mode vacuum, an arbitrary passive quadratic addition leaves $v_{FS}$ exactly unchanged while the instability spectrum can change; in the detuned-amplifier family below the variation is continuous and carries $\lambda_{max}$ from the bound of Theorem 2 down to zero at rigorously fixed $v_{FS}$.

The separation is one-sided, and the surviving direction is sharp.

*Theorem 2.* For every number of bosonic modes,

$$\lambda_{\max} \leq \sqrt{2}\, v_{\mathrm{FS}} \tag{3}$$

and the constant √2 is sharp for every $N$.

*Proof.* By Bendixson's theorem the real part of every eigenvalue of $M$ is bounded by the largest eigenvalue $\mu_{max}(S)$ of its symmetric part. With $M = JG$ and $G$ symmetric, $S = (JG - GJ)/2$, and $J^2 = -1$ gives $SJ + JS = 0$: the symmetric part of a Hamiltonian flow anticommutes with the symplectic form. Its spectrum is therefore symmetric about zero, since $Sx = \mu x$ implies $S(Jx) = -\mu(Jx)$, so the eigenvalues occur in opposite pairs and $\|S\|_F^2 \geq 2\mu_{max}^2(S)$. Hence $\lambda_{max} \leq \mu_{max}(S) \leq \|S\|_F/\sqrt{2} = \sqrt{2}\, v_{FS}$ by Theorem 1. Equality holds for a resonant single-mode pure squeezer, $G = \mathrm{diag}(\varepsilon, -\varepsilon)$, which has $\lambda_{max} = |\varepsilon|$ and $v_{FS} = |\varepsilon|/\sqrt{2}$; adjoining spectator modes with vanishing generator extends the saturating example to every $N$. ■

The Bendixson step is the standard numerical-abscissa bound [9]. Symplectic pairing of the stretching rates converts it to the sharp geometric bound. Section 5 gives the corresponding identity for the full unstable spectrum and characterizes equality.

**4. Reciprocal families in a parametric amplifier**

The realizable single-mode form is a parametrically pumped mode in the frame rotating at half the pump frequency,

$$\hat{H} = \delta(\hat{x}^2 + \hat{p}^2)/2 + \varepsilon(\hat{x}^2 - \hat{p}^2)/2 \tag{4}$$

an ideal single-mode model of a Josephson parametric amplifier. Here $\delta$ is the detuning, $\varepsilon$ is the pump coefficient, and $G = \mathrm{diag}(\delta + \varepsilon, \delta - \varepsilon)$. The flow is hyperbolic for $|\varepsilon| > |\delta|$, with exponent $\lambda = \sqrt{\varepsilon^2 - \delta^2}$. At $|\varepsilon| = |\delta|$ the spectral abscissa vanishes. Equation (2) gives

$$v_{FS} = |\varepsilon|/\sqrt{2} \tag{5}$$

for every detuning. The detuning term is passive, so it leaves the symmetric part of the flow unchanged in the bare-vacuum metric, as in Corollary 2.

Two one-parameter families follow. *Family A*: vary $\delta$ holding $\varepsilon^2 - \delta^2 = \lambda_0^2$, so that $\lambda = \lambda_0$ at every point while $\sqrt{2}\, v_{FS} = \sqrt{(\lambda_0^2 + \delta^2)}$ grows, reaching 1.414, 2.236 and 3.162 times its on-resonance value at $\delta/\lambda_0 = 1, 2, 3$ [Fig. 1(a)]. *Family B*: vary $\delta$ at fixed $\varepsilon$, so that $v_{FS}$ is constant while $\lambda = \varepsilon\sqrt{(1 - (\delta/\varepsilon)^2)}$ falls from $\varepsilon$ to zero [Fig. 1(b)]. Inverted, the ratio is the saturation statement of Theorem 2 for this device,

$$\lambda/(\sqrt{2}\, v_{FS}) = \sqrt{1 - \delta^2/\varepsilon^2} \leq 1 \tag{6}$$

equal to unity on resonance, where the generator is a pure squeezer, and falling below it as detuning adds passive rotation that $v_{FS}$ does not register.

The pair is the content of the theorem in laboratory terms: same instability $\nRightarrow$ same geometric speed, and same geometric speed $\nRightarrow$ same instability. The detuning of a parametric amplifier is directly tunable and independently measurable, and the pump amplitude is calibrated independently of the geometry, so the two families predict not only that the bound of Eq. (3) is saturated at $\delta = 0$ but exactly how the two rates separate, in both directions, as the pump is detuned. Because the generator is indefinite above threshold and has no ground state, the speed must be referred to a stated initial condition, here the bare-mode vacuum before the pump is switched on; Eq. (5) is independently confirmed by direct Fock-basis evaluation [29].

The separation is between measurable trajectories. Evolving the bare-mode vacuum under Eq. (4) gives the mean occupation exactly,

$$\langle n(t)\rangle \;=\; 2(v_{\mathrm{FS}}/\lambda)^2\,\sinh^2(\lambda t) \tag{7}$$

which assigns the two rates different roles: the exponent sets the rate of the exponential, the geometric speed sets its amplitude. The closed form is verified against direct covariance propagation [29]. At short times $\langle n\rangle \to 2v_{\mathrm{FS}}^2t^2$ independent of $\lambda$, and, at the threshold edge $\lambda \to 0$, that form is exact at all times; at long times $\langle n\rangle \to (v_{\mathrm{FS}}^2/2\lambda^2)e^{2\lambda t}$. Along Family A the amplitude $1 + \delta^2/\lambda_0^2$ takes the values 1, 2, 5 and 10 at $\delta/\lambda_0 = 0, 1, 2, 3$ while the asymptotic rate is common to six digits [Fig. 1(c)]; along Family B the short-time behaviour is common while the slopes fan out [Fig. 1(d)]. The occupation at $\varepsilon t = 8$ spans more than four orders of magnitude (Table I). Family A members also differ in the shape of the covariance: at $\lambda_0 t = 0.8$ the quadrature anisotropy across the four members runs 24.5, 67.0, 313.9 and 1123.6 while the squeezing axis rotates from 135.0° to 103.3° [29]. Holding $\lambda$ fixed therefore does not leave a reparametrization of one flow.

Table I gives both quantities along the two families.

The two families in Fig. 1 compare unstable configurations. Family B reaches zero exponent only at its marginal endpoint. Below threshold the bare-vacuum speed remains nonzero while the flow is elliptic.

TABLE I. Closed-system dynamics along the reciprocal families, evaluated from the bare-mode vacuum. Rates and times are in units of $\lambda_0$ for Family A and of $\varepsilon$ for Family B; the fidelity decay is evaluated at 0.02 in those units, well inside the quadratic regime. In Family A the exponent is fixed and the short-time decay varies; in Family B the short-time decay is fixed and the long-time occupation varies.

| Family | detuning | $1 - F$ | long-time quantity |
|---|---|---|---|
| A | $\delta/\lambda_0 = 0$ | $2.000\times10^{-4}$ | $\mathrm{d}\ln\langle n\rangle/\mathrm{d}t = 2.000000$ |
| A | 1 | $3.998\times10^{-4}$ | 2.000000 |
| A | 2 | $9.986\times10^{-4}$ | 2.000000 |
| A | 3 | $1.994\times10^{-3}$ | 2.000000 |
| B | $\delta/\varepsilon = 0$ | $2.000\times10^{-4}$ | $\langle n\rangle$ at $\varepsilon t = 8$: $2.22\times10^{6}$ |
| B | 0.5 | $2.000\times10^{-4}$ | $3.47\times10^{5}$ |
| B | 0.9 | $1.999\times10^{-4}$ | $1.40\times10^{3}$ |
| B | 0.99 | $1.999\times10^{-4}$ | $9.62\times10^{1}$ |

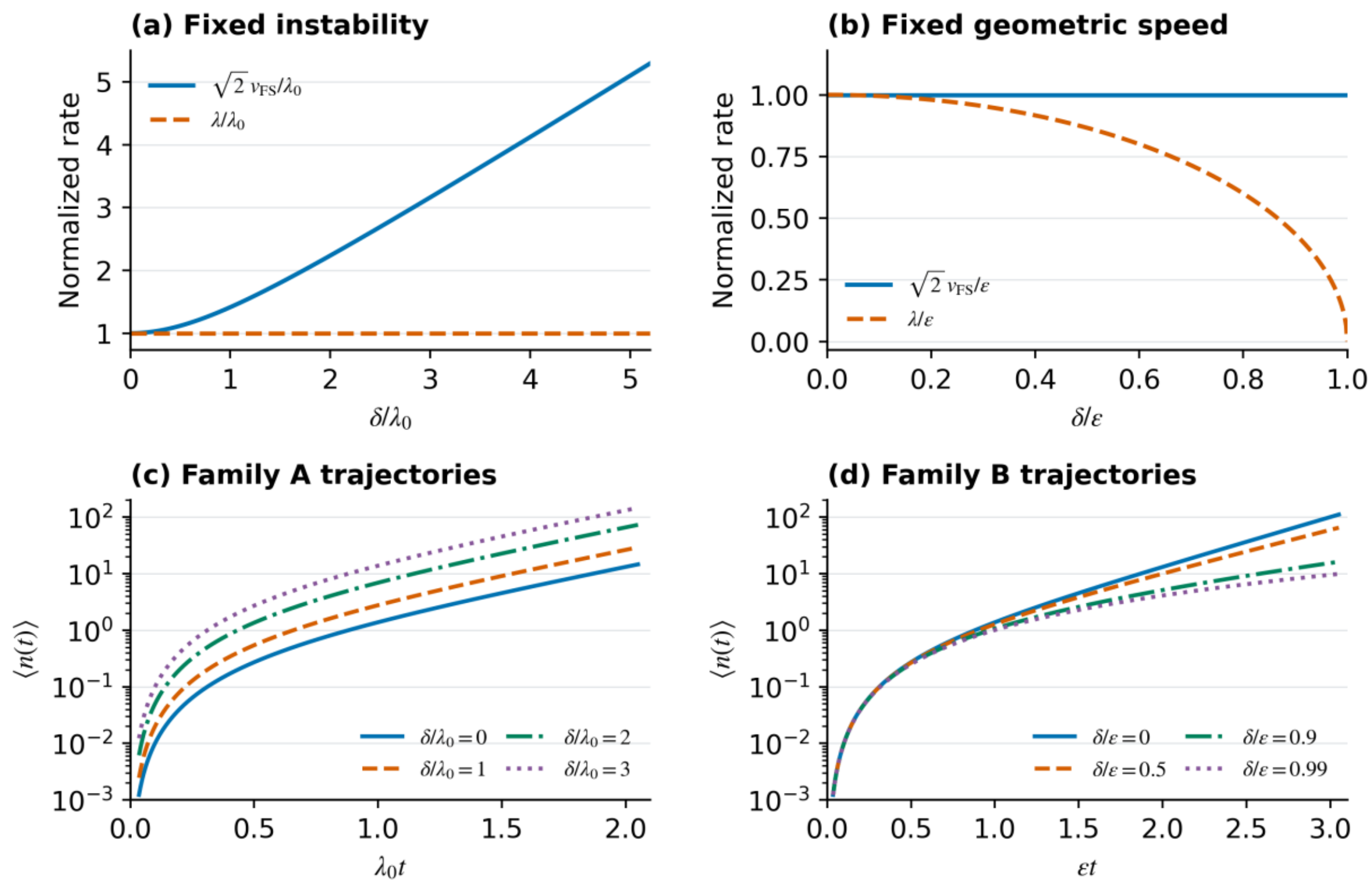

FIG. 1. Reciprocal separation in a detuned parametric amplifier. (a) Family A: $\varepsilon$ and $\delta$ co-varied so that $\lambda = \lambda_0$ is fixed exactly (dashed) while the geometric rate (solid) grows. (b) Family B: $\varepsilon$ fixed, so the geometric rate is constant (solid) while $\lambda/\varepsilon$ (dashed) falls to zero at threshold; by Eq. (6) the dashed curve is the approach to the bound of Theorem 2. (c) Mean occupation of Eq. (7) for Family A: common exponential rate, amplitudes differing by an order of magnitude. (d) The same for Family B: common short-time behaviour, fanning slopes. Family A requires $\varepsilon$ and $\delta$ to be co-varied along a hyperbola; Family B varies $\delta$ at fixed $\varepsilon$.

## 5. Optimal Gaussian inputs for a fixed generator

Every centered pure Gaussian input has covariance $\Sigma = VV^{\mathrm{T}}/2$, with $V$ real symplectic. The generator in quadratures adapted to that input is $M_V = V^{-1}MV$. The freedom to multiply $V$ on the right by an orthogonal symplectic matrix leaves the covariance unchanged and acts by orthogonal similarity on $M_V$. Frobenius norms in these adapted quadratures therefore give unambiguous quantities for the chosen state.

**Theorem 3.** Define $a_j = \mathrm{Re}\,\mu_j > 0$ over the eigenvalues of $M$ with positive real part, counted algebraically. Let $D_{\mathrm{H}}(X)$ denote the Henrici departure from normality [30], with $D_{\mathrm{H}}(X)^2 = \|X\|_{\mathrm{F}}^2 - \sum|\mu_j(X)|^2$. Then for every centered pure Gaussian input,

$$2v_{\mathrm{FS}}(\Sigma)^2 = \sum_j a_j^2 + D_{\mathrm{H}}(M_V)^2/4 \tag{8}$$

and the optimal value at fixed Hamiltonian is

$$\inf_{\Sigma} v_{\mathrm{FS}}(\Sigma) = (\sum_j a_j^2/2)^{1/2} \tag{9}$$

where the infimum runs over centered pure Gaussian inputs. An input attaining Eq. (9) exists if and only if $M$ is diagonalizable over $\mathbb{C}$. In that case the optimal covariances are precisely those for which $M_V$ is normal.

**Proof.** Equation (2) applies in the adapted quadratures and gives $8v_{\mathrm{FS}}(\Sigma)^2 = \|M_V\|_{\mathrm{F}}^2 + \mathrm{Tr}(M_V^2)$. By the definition of $D_{\mathrm{H}}$ and trace invariance, the right side is $D_{\mathrm{H}}(M_V)^2$ plus $\sum[|\mu_j|^2 + \mathrm{Re}(\mu_j^2)]$. Each summand in brackets is $2(\mathrm{Re}\,\mu_j)^2$. The Hamiltonian pairing $\mu \leftrightarrow -\mu$ yields Eq. (8). By the complex Schur decomposition, $D_{\mathrm{H}}$ is the Frobenius norm of the strictly upper-triangular part and vanishes exactly at normality.

For the infimum, minimize $\|X\|_{\mathrm{F}}^2$ over the Euclidean closure of the real symplectic similarity orbit of $M$. A minimizer exists because the norm is coercive and the closure is nonempty and closed. The closure stays inside the real Hamiltonian Lie algebra and preserves the characteristic polynomial. It is also invariant under symplectic conjugation. At a minimizer $X$, $K = [X, X^{\mathrm{T}}]$ belongs to that Lie algebra, which is closed under transpose and commutators. The derivative at $t = 0$ of $\|\exp(tK)\, X \exp(-tK)\|_{\mathrm{F}}^2$ is $2\|K\|_{\mathrm{F}}^2$, hence $K = 0$. The normal minimizer lies in the closure and can be approached from the orbit. Equation (8) therefore gives Eq. (9).

Attainment requires a normal representative in the orbit and thus complex diagonalizability. Conversely, the symplectic canonical form for a diagonalizable generator has only semisimple blocks [19]. Real opposite eigenvalues give diagonal blocks, imaginary pairs give skew-symmetric rotation blocks, and a complex quartet gives a block $\mathrm{diag}(B, -B^{\mathrm{T}})$ with $B$ a real scalar-plus-rotation matrix. The zero eigenspace contributes a zero block. This real symplectic representative is normal and attains Eq. (9).

At a fixed input, Eq. (8) gives the full-spectrum inequality. For $v_{\mathrm{FS}} > 0$, equality in the largest-exponent bound of Eq. (3) holds exactly when $M_V$ is normal and has one eigenvalue with positive real part, counted with multiplicity. If all the real parts vanish and $M$ is defective, the speed has zero infimum but is positive for every finite Gaussian preparation.

### 5.1. Optimal speed at finite squeezing

Return to Eq. (4), taking $\delta \geq 0$ and $\varepsilon > 0$ for the calculation. The optimal speed depends only on their magnitudes. Write $x = |\delta/\varepsilon|$, $v_{\mathrm{bare}} = |\varepsilon|/\sqrt{2}$, and let $R \geq 0$ be the maximum input squeezing. A centered pure Gaussian input with squeeze $r \leq R$ has covariance $\Sigma = Q\,\mathrm{diag}(e^{-2r}, e^{2r})Q^{\mathrm{T}}/2$, with $Q$ a rotation by $\theta$. In one mode the Gaussian variance identity in Sec. 3 reduces to $2v_{\mathrm{FS}}^2 = [\mathrm{Tr}(G\Sigma)]^2 - \det G$, since $\det \Sigma = 1/4$. The state-dependent trace is

$$\mathrm{Tr}(G\Sigma) = \delta \cosh(2r) - \varepsilon \cos(2\theta) \sinh(2r) \tag{10}$$

For fixed $r$, varying $\theta$ sweeps an interval centered at $\delta \cosh(2r)$ with half-width $\varepsilon \sinh(2r)$. Minimizing the absolute trace first over that interval and then over $0 \leq r \leq R$ gives three branches for $R > 0$. For $x \leq \tanh(2R)$, the optimum reaches the invariant hyperbolic floor. For $x \geq \coth(2R)$, a zero-speed Gaussian input is accessible. Between these boundaries the optimum uses $r = R$ along the appropriate quadrature. The resulting normalized speed is

$$v_{\mathrm{opt}}/v_{\mathrm{bare}} = \sqrt{1 - x^2}, \quad 0 \leq x \leq \tanh(2R) \tag{11a}$$

$$v_{\mathrm{opt}}/v_{\mathrm{bare}} = \cosh(2R) - x \sinh(2R), \quad \tanh(2R) \leq x \leq \coth(2R) \tag{11b}$$

$$v_{\mathrm{opt}}/v_{\mathrm{bare}} = 0, \quad x \geq \coth(2R) \tag{11c}$$

At $R = 0$ the only centered input is the bare vacuum, so the ratio is unity. At the nonzero Hamiltonian threshold $x = 1$, the flow is nilpotent and Eq. (11b) gives the exact preparation cost. The photon budget $\bar{n} = \sinh^2 R$ refers to the mean bare-mode occupation of the input, before evolution under Eq. (4).

$$v_{\mathrm{opt}}/v_{\mathrm{bare}} = e^{-2R} = (\sqrt{\bar{n}+1} - \sqrt{\bar{n}})^2, \quad x = 1 \tag{12}$$

Figure 2 shows the squeezing-limited interval around the threshold and the residual speed there. To test Eq. (12), prepare centered squeezed inputs with their squeezed axis along the quadrature that enters the threshold Hamiltonian. Hold the pump and detuning fixed across preparations. Determine $v_{\mathrm{FS}}^2$ from the short-time curvature of the squared fidelity to each input. In the unitary Gaussian model the overlap can be obtained from input and output covariances as $F(t) = 1/\sqrt{\det(\Sigma(0) + \Sigma(t))}$. At $\delta = \varepsilon > 0$, the threshold Hamiltonian is $\hat{\mathrm{H}} = \varepsilon\hat{x}^2$. A Gaussian integral over the aligned input quadrature gives the exact squared fidelity

$$F_R(t) = [1 + (\varepsilon t\, e^{-2R})^2]^{-1/2} \tag{13}$$

for the unitary threshold protocol. With single-photon loss from the bare mode, a squeezed input acquires the initial linear fidelity loss $\kappa \sinh^2(R)t$. Extracting the unitary curvature in a lossy device therefore requires modeling the dissipative contribution or extrapolating to zero loss. Appendix A treats the distinct lossy threshold for the bare-vacuum protocol.

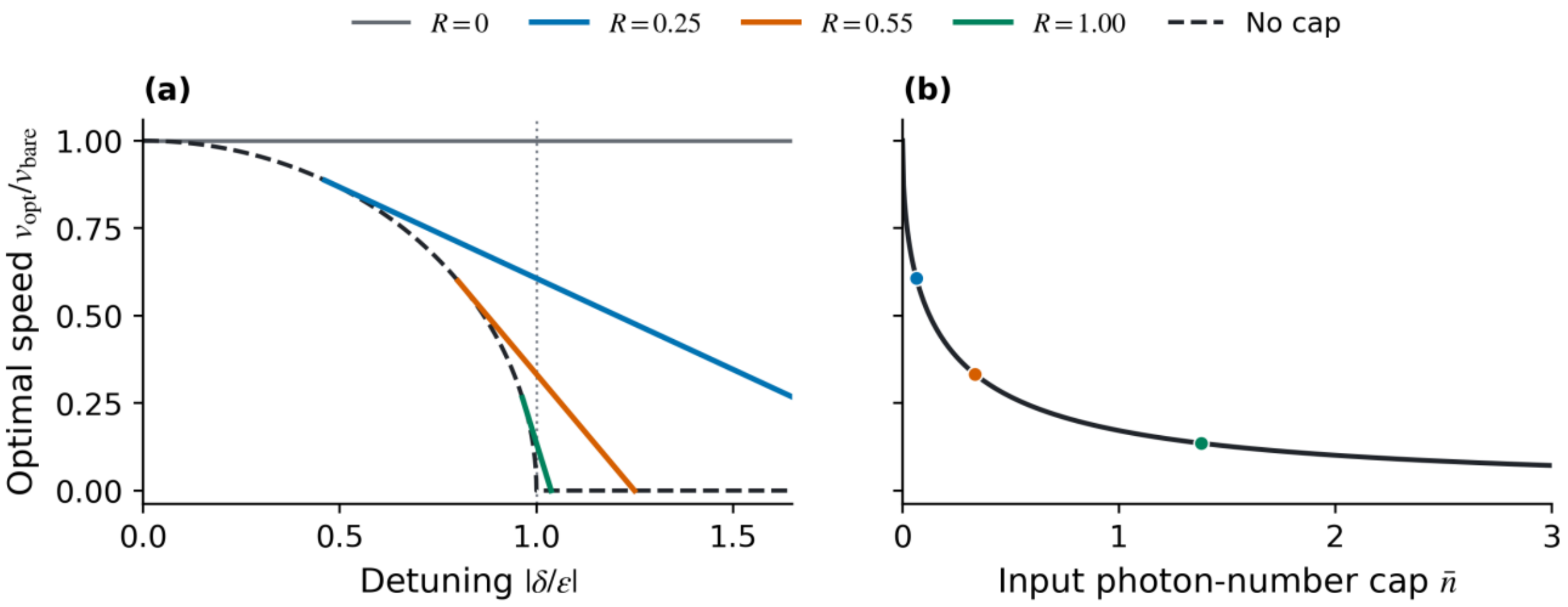

**FIG. 2.** Optimal initial speed in the ideal one-mode amplifier. (a) Normalized speed versus $x = |\delta/\varepsilon|$ for finite input squeezing caps and for the unconstrained Gaussian infimum (dashed). The dotted line marks $x = 1$. Coincident branches share the dashed trace. (b) Exact threshold law as a function of the maximum mean input photon number $\bar{n} = \sinh^2 R$. The normalized speed remains positive for every finite budget.

### 5.2. Growth threshold from two return-fidelity traces

We now use the same amplifier to ask an inverse question. Assume a constant Hamiltonian as in Eq. (4), centered Gaussian inputs, and a common phase-insensitive Markovian channel. Let $\kappa \geq 0$ be the energy damping rate and $d \geq \kappa/2$ the isotropic quadrature diffusion coefficient in the convention $\Sigma_0 = I/2$. For a thermal bath with $\kappa > 0$, $d = \kappa(n_{th} + 1/2)$. Both bath parameters are unknown. The covariance evolves as

$$\Sigma = A\Sigma + \Sigma A^{\mathrm{T}} + dI, \quad A = M - \kappa I/2 \tag{14}$$

where $M_{12} = \delta - \varepsilon$, $M_{21} = -\delta - \varepsilon$, and the diagonal of $M$ vanishes. Use two preparations, the bare vacuum and a squeezed vacuum of known squeeze $r > 0$, rotated by $\pi/4$ from the quadratures that diagonalize the quadratic form in Eq. (4). With $c = \cosh(2r)$, $s = \sinh(2r)$ and $P(x,p)^{\mathrm{T}} = (p,x)^{\mathrm{T}}$, their covariances are $\Sigma_0 = I/2$ and $\Sigma_{45} = (cI - sP)/2$. The same Hamiltonian and bath act in both runs.

Let $F_i(t) = \mathrm{Tr}(\rho_i(0)\rho_i(t))$ be the squared return fidelity, with $i = 0{,}45$. Both initial states are pure and centered. Hence $F_i(t)^{-2} = \det(\Sigma_i + \Sigma_i(t))$, with the expansion

$$F_i(t)^{-2} = 1 + \ell_i t + b_i t^2 + O(t^3) \tag{15}$$

**Theorem 4.** The coefficients of the two traces determine $\kappa$, $d$, $\varepsilon^2$ and $\delta^2$. In the stated normalization they satisfy

$$\ell_0 = 2d - \kappa, \quad \ell_{45} = 2dc - \kappa \tag{16a}$$

$$b_0 = \varepsilon^2 + 3\kappa^2/4 - 2\kappa d + d^2 \tag{16b}$$

$$b_{45} - b_0 = \delta^2 s^2 - \kappa(\ell_{45} - \ell_0) \tag{16c}$$

Since $r > 0$, the inverse is

$$d = (\ell_{45} - \ell_0)/[2(c-1)], \quad \kappa = 2d - \ell_0 \tag{17a}$$

$$\varepsilon^2 = b_0 - 3\kappa^2/4 + 2\kappa d - d^2 \tag{17b}$$

$$\delta^2 = [b_{45} - b_0 + \kappa(\ell_{45} - \ell_0)]/s^2 \tag{17c}$$

The largest real drift exponent of Eq. (14) is positive exactly when the following growth margin is positive:

$$m = \varepsilon^2 - \delta^2 - \kappa^2/4 \tag{18}$$

**Proof.** Set $X = \Sigma_i$, $W = (2X)^{-1}$ and $H(Y) = MY + YM^{\mathrm{T}}$. The first two covariance derivatives are $U = H(X) - \kappa X + dI$ and $Z = H(U) - \kappa U$. Expanding the determinant of $2X + tU + t^2 Z/2$, with $\det(2X) = 1$, gives $\ell_i = \mathrm{Tr}(WU)$ and $2b_i = \mathrm{Tr}(WZ) + [\mathrm{Tr}(WU)]^2 - \mathrm{Tr}[(WU)^2]$. The Hamiltonian contribution to the slope vanishes because $\mathrm{Tr}(WH(X)) = \mathrm{Tr}\, M = 0$. At quadratic order the mixed term proportional to $\kappa$ vanishes for the same reason. The two terms proportional to $d$ cancel by $2\mathrm{Tr}(W^2 H(X)) = \mathrm{Tr}[W(M + M^{\mathrm{T}})]$, using $WX = I/2$.

For pure damping and diffusion the determinant expansion gives $b_i = 3\kappa^2/4 - 2\kappa d c_i + d^2$, with $c_0 = 1$ and $c_{45} = c$. The purely Hamiltonian contribution is $2v_{\mathrm{FS}}^2$ from Eq. (2) in the adapted quadratures. The one-mode variance formula in Sec. 5.1 gives $2v_{\mathrm{FS}}^2 = \varepsilon^2$ for $\Sigma_0$ and $2v_{\mathrm{FS}}^2 = \varepsilon^2 + \delta^2 s^2$ for $\Sigma_{45}$, since $\mathrm{Tr}(G\Sigma_0) = \delta$ and $\mathrm{Tr}(G\Sigma_{45}) = \delta c$. This gives Eqs. (16a)–(16c). The eigenvalues of the drift are $-\kappa/2 \pm \sqrt{\varepsilon^2 - \delta^2}$, with the square root taken on the complex plane, which gives Eq. (18).

For a bare-vacuum return, both initial coefficients in Eqs. (16a) and (16b) are independent of detuning. Thus at fixed pump and bath they are identical for amplifiers on opposite sides of the growth boundary. Higher orders can carry detuning information. For example, in the unitary vacuum trace the first detuning-dependent term occurs at order $t^4$. The two preparations resolve this ambiguity but retain an exact symmetry of the data. The swap $P$ conjugates $M(\delta)$ to $M(-\delta)$ and leaves both initial covariances and the diffusion invariant. The two fidelities are therefore even in detuning at all times. The unsigned detuning suffices for the stability margin.

The return may be measured by undoing the preparation and projecting onto the vacuum of the same mode, or inferred from homodyne covariance measurements of the centered input and output. The latter yield additional information. The theorem specifies what the scalar returns alone determine. Repeated preparations at calibrated times are needed to estimate the initial coefficients. The inference assumes a calibrated squeeze and angle and common isotropic drift loss and diffusion. For small squeezing the denominators $c-1$ and $s^2$ make the inference sensitive to readout errors.

An angle error gives a separate preparation systematic. If the squeeze angle is $\theta$ but the inversion assumes $\pi/4$, Eq. (10) gives the bias $\delta_{est}^2-\delta^2=\varepsilon^2\cos^2(2\theta)-2\varepsilon\delta(c/s)\cos(2\theta)$ for exact initial coefficients. At the benchmark boundary below, a $0.1°$ angle offset changes $m/\varepsilon^2$ by approximately $0.0096$ in magnitude. A perpendicular squeezed control cancels the term linear in the angle error when the two quadratic coefficients are averaged. The two-input inversion as stated uses a calibrated angle.

Figure 3 uses direct affine covariance propagation of Eq. (14). In the simulation $\kappa/\varepsilon=0.4$, $2d/\kappa=1.5$ and $r=0.45$. The normalization by the true pump coefficient is only for presenting the benchmark. The inverse in Eqs. (17a)–(17c) uses physical time and no supplied pump or bath rates. For each trace we fit the three powers $t$, $t^2$ and $t^3$ to eight equally spaced positive times ending at $\varepsilon t=0.10$, with intercept fixed at one. The cubic term absorbs the leading correction to the quadratic law.

To quantify finite-data sensitivity, let each sampled value of $F_i^{-2}$ have an additive readout error of magnitude at most $\eta$. The coefficient error bounds $E_\ell$ and $E_b$ are $\eta$ times the sums of absolute weights in the first and second rows of the design pseudoinverse. Write $L=\ell_{45}-\ell_0$ and $g=\kappa-d$. The bounds for $L$, $\kappa$ and $g$ are $E_L=2E_\ell$, $E_\kappa=E_L/(c-1)+E_\ell$, and $E_g=E_L/[2(c-1)]+E_\ell$, respectively. Substituting Eqs. (17a)–(17c) into Eq. (18) gives the margin in the form

$$m=(1+1/s^2)b_0-b_{45}/s^2-g^2-\kappa L/s^2 \tag{19}$$

The magnitude of its change under the stated readout errors is at most

$$(1+2/s^2)E_b+(2|g|+E_g)E_g+(|\kappa|E_L+|L|E_\kappa+E_\kappa E_L)/s^2 \tag{20}$$

with $g$, $\kappa$ and $L$ evaluated at the unperturbed fit for the displayed envelope. This deterministic bound describes only the response of the finite-window estimator to the chosen readout errors.

For $\eta=10^{-6}$, the readout-error half-width in Fig. 3(b) is about $0.0131\varepsilon^2$. The difference between the noiseless fit and the exact margin is at most $0.00101\varepsilon^2$ on the 200-point simulation grid. The cases $\delta/\varepsilon = 0.955$ and $0.995$ give exact margins of $0.047975\varepsilon^2$ and $-0.030025\varepsilon^2$, respectively, even though their bare-vacuum slopes and curvatures coincide. At the dissipative boundary the envelope straddles zero. The benchmark is a calculation for specified readout and preparation conditions.

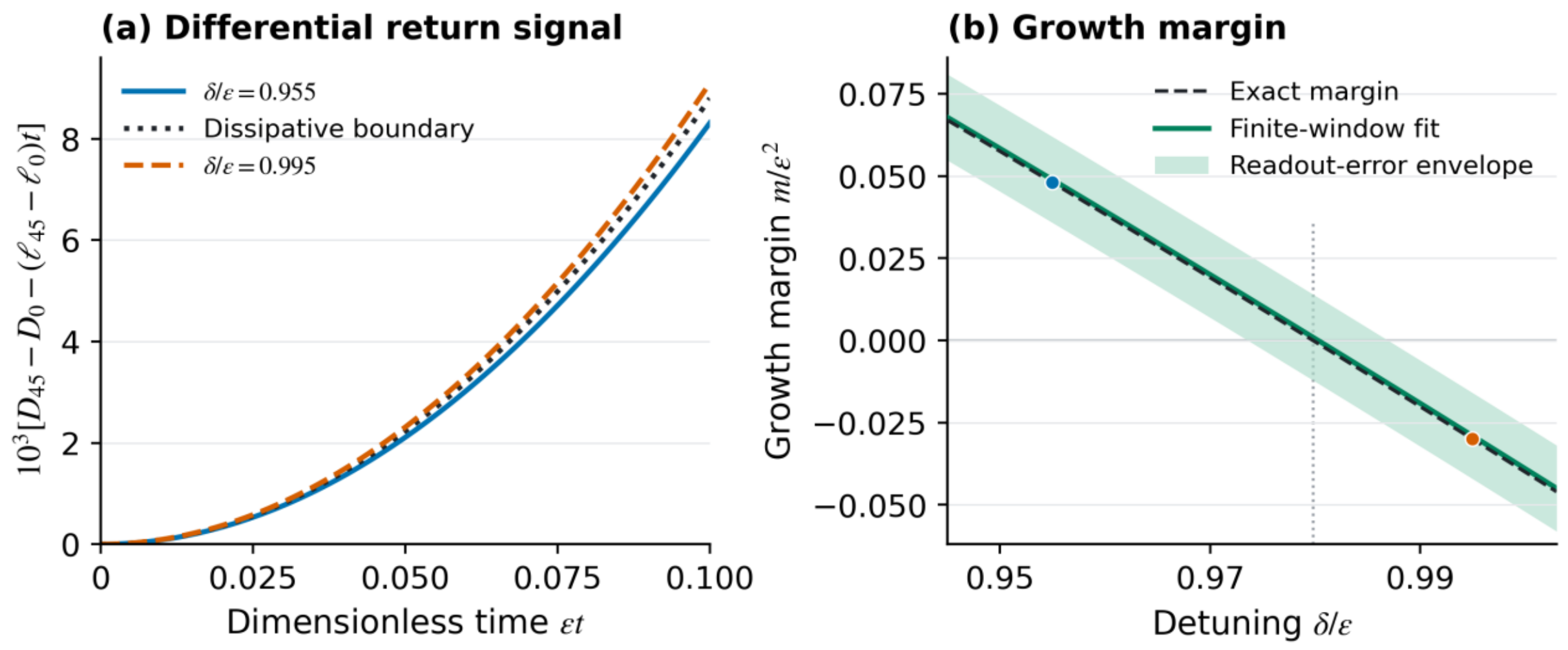

**FIG. 3.** Growth-threshold reconstruction from two return fidelities. (a) Difference of the determinant traces $D_i = F_i^{-2}$ after subtracting the linear term of the model, for two detunings and the dissipative boundary. (b) Exact margin and noiseless finite-window fit. The shaded envelope bounds the change of that fit for errors of at most $10^{-6}$ per determinant sample. The separation of the exact and fitted curves shows the finite-window bias. The markers identify the two detunings in panel (a).

### 6. Collective-spin example

The local quadratic generator of the twisting-and-turning collective-spin Hamiltonian studied in Ref. [31] has $G = \mathrm{diag}(-\Omega, \tilde{\chi} - \Omega)$ near its saddle. For $\tilde{\chi} > 0$, Eq. (2) gives $v_{\mathrm{FS}} = \tilde{\chi}/(2\sqrt{2})$, independent of the control ratio $u = \Omega/\tilde{\chi}$. The unstable exponent $\Lambda = \tilde{\chi}\sqrt{u(1-u)}$ varies with $u$ and peaks at $u = 1/2$. The control changes only the passive part of the generator, explaining the flat speed. Ref. [31] considers a quantum speed limit for a squeezed-state preparation protocol, while the present speed uses a fixed vacuum input. In the related spin model of Ref. [17], two local curvatures determine a short-time gain scale distinct from the exponent. In a single-mode quadratic reduction that scale corresponds to the largest eigenvalue of the symmetric part of the flow, $\mu_{\max}(S) = \sqrt{2}\, v_{\mathrm{FS}}$.

## 7. Limits of frozen spectral diagnostics

The spectral results in Secs. 3 and 5 concern a fixed generator. For a time-dependent generator $M(t)$ the same algebra applies to a frozen matrix and a specified reference. Its spectral abscissa generally differs from the growth rate of the time-ordered propagator.

In frequency-normalized moving quadratures for the chirped mode of Ref. [29], set $a = \dot{\Omega}/(2\Omega)$. The symmetric stretching rate is $|a|$, and the frozen spectral abscissa is $\sqrt{max(a^2 - \Omega^2, 0)}$. Their squared difference is $\min(a^2, \Omega^2)$, equal to $\Omega^2$ on the hyperbolic branch. In the fixed laboratory quadratures the positive-frequency oscillator instead has an elliptic frozen generator throughout.

At parametric resonance, with $\Omega(t) = \Omega_0[1 + m\cos(2\Omega_0 t)]$ and $0 < m < 1$, the oscillator has an elliptic frozen generator in laboratory quadratures at every instant. The Floquet exponent approaches $m\Omega_0/2$ at weak modulation [29]. The positive rate belongs to the monodromy over a period. A local rate along the instantaneous eigenvacuum path and the speed generated by the Hamiltonian on a fixed input are different observables.

Equation (3) bounds the autonomous spectral abscissa. For a driven system with periodic parameters, a bound on Floquet growth would instead have to involve data over a period. The mixed-state Bures speed and dynamics with loss require their own treatment. Appendix A addresses the spectral shift and a finite observation window for a vacuum input with uniform loss.

## 8. Discussion

The metric in Ref. [32], Eqs. (10)–(11), gives the arc length $L_{\mathrm{FS}} = |\ln(\Omega_f/\Omega_i)|/\sqrt{8}$ by integration along a monotonic one-mode instantaneous-vacuum path. Ref. [33] introduces the gap-normalized ratio $A_n = v_{\mathrm{FS}}/\Delta_n$ and obtains $A_0 = \eta/\sqrt{8}$ in the convention $\Delta_0 = \Omega$. Ref. [34] proposes a realization-dependent comparison for tangent dynamics in driven nonlinear systems. The present spectral identity concerns the unitary speed of a specified input under the same fixed quadratic generator whose spectrum determines the instability.

In Family A the one-mode spectrum stays fixed as the speed changes. In Family B the spectrum changes at fixed bare-vacuum speed. The variational result addresses a third control: fixing the Hamiltonian and changing the input covariance. It identifies the part of the speed that persists after this control has been optimized and quantifies the effect of a finite preparation budget at the threshold.

The relation to non-normality is quantitative. At each Gaussian input the difference between the squared speed and the squared spectral floor is $D_{\mathrm{H}}(M_V)^2/8$. Earlier work relates non-normality to transient and asymptotic growth in quadratic bosonic systems [12–16,18]. Its contribution here is to the initial quantum distinguishability speed. Equation (9) is the infinitesimal Fubini–Study counterpart of the Siegel translation-length formula in Ref. [21]. In that normalization the right side of Eq. (9) equals $\ell_S(\exp(tM))/(2\sqrt{2}\,|t|)$ for $t \neq 0$. The fixed-input identity and the finite-squeezing calculation connect that geometry to energy variance and a preparation-dependent fidelity prediction.

Parametric-resonator detuning has been studied through critical sensing and frequency-resolved output Fisher information [35–36]. Generic Gaussian process tomography with coherent probes uses richer output data [37]. Phase-resolved projective derivatives also permit local reconstruction of sparse time-dependent bosonic Hamiltonians [38]. Here four scalar Taylor coefficients of two return probabilities suffice for a one-mode generator with unknown common loss and diffusion. They locate the linear growth boundary without requiring stationary output. Statistical precision depends on the readout and preparation errors.

## 9. Conclusion

For an autonomous quadratic bosonic Hamiltonian, the entire unstable spectrum fixes the lowest attainable limit of the Fubini–Study speed over centered pure Gaussian inputs. A finite input attains the limit precisely for a diagonalizable flow generator. For a specified input, the excess squared speed is the squared Henrici departure in its adapted quadratures divided by eight. The sharp bare-vacuum bound on the largest exponent follows as a corollary.

For the one-mode amplifier, finite squeezing sets a quantitative lower limit on the speed at the Hamiltonian threshold. Two short-time return traces also suffice to infer the open-system growth threshold under a common isotropic Gaussian bath. This separates preparation-dependent distinguishability from the drift stability that can be recovered by changing the preparation.

**Appendix A. Experimental window in a lossy resonator**

Uniform single-photon loss into a vacuum bath changes the quadrature drift to $M_{\text{open}} = M - \kappa I/2$, with exponents $-\kappa/2 \pm \lambda$ in the closed-system hyperbolic sector. The bare input vacuum is dark under this loss. Its initial quadratic fidelity coefficient is the same as in the closed model, and the finite-time estimate below uses $\kappa t \ll 1$. Net instability requires $\lambda > \kappa/2$. On resonance the pump threshold is $\varepsilon_{th} = \kappa/2$. At fixed $\varepsilon > \varepsilon_{th}$, Family B remains unstable for $|\delta/\varepsilon| < \sqrt{1 - (\kappa/2\varepsilon)^2}$. Its closed-system exponent decreases from $\varepsilon$ to $\kappa/2$, giving the ratio $h = \varepsilon/\varepsilon_{th}$. Family A instead holds $\lambda = \lambda_0$ and its speed range is set by the maximum pump $\varepsilon_{\max}$ for which the linear model is accurate [39]. For $\lambda_0 = \kappa$ and $\varepsilon_{\max} = 5\varepsilon_{th}$, the speed in Family A varies by a factor of 2.5, giving a factor of 6.25 in the quadratic fidelity coefficient.

Table II evaluates Family B for parameters representative of resonator-based Josephson parametric amplifiers, $\kappa/2\pi = 20$ MHz and $\varepsilon/2\pi = 50$ MHz — the linewidth is of the order of the 30 MHz squeezing bandwidth reported in Ref. [40] — so that $v_{\text{FS}}/2\pi = 35.36$ MHz identically at every row. The two observables are read in separated time windows. The quadratic fidelity decay must be taken at $\kappa t \ll 1$: at $t = 0.5$ ns, $\kappa t = 0.063$ and $1 - F = 0.012$. The exponent is read once the decaying branch of the flow has died away, on the scale $1/(\lambda + \kappa/2) = 2.65$ ns on resonance, and before the occupation reaches the Kerr saturation scale $n_{\text{sat}} \approx 100$. Integrating the covariance equation, the logarithmic growth rate settles to within 5 % of $2(\lambda - \kappa/2)$ after 7.1 ns on resonance and 13.7 ns at $\delta/\varepsilon = 0.8$, against $t_{\text{sat}} = 11.5$ and 18.2 ns, leaving 2.2 and 1.1 e-folds of clean exponential growth. The window narrows as the pump is detuned further, to 0.4 e-folds at $\delta/\varepsilon = 0.9$, and at $\delta/\varepsilon = 0.95$ the rate has not settled before saturation. Table II is therefore a linear-model feasibility estimate rather than a claim that the asymptotic exponent can be extracted at every listed detuning. The resonant end of Family B is where both measurements fit inside the linear regime, and there they are separated by more than an order of magnitude in $t$: the fidelity decay is read at 0.5 ns and the exponent between 7 and 11.5 ns.

TABLE II. Family B in a lossy resonator, $\kappa/2\pi = 20$ MHz and $\varepsilon/2\pi = 50$ MHz, i.e. $\varepsilon/\varepsilon_{\mathrm{th}} = 5$ with $\varepsilon_{\mathrm{th}} = \kappa/2$, for which $\nu_{\mathrm{FS}}/2\pi = 35.36$ MHz at every row. The occupation is obtained by integrating $\mathrm{d}\Sigma/\mathrm{d}t = M_{\mathrm{open}}\Sigma + \Sigma M_{\mathrm{open}}^{\mathrm{T}} + \kappa\Sigma_{\mathrm{vac}}$ from vacuum, so the vacuum-noise refill accompanying the loss is included; $t_{\mathrm{sat}}$ is the time to reach 100 photons.

| $\delta/\varepsilon$ | $\delta/2\pi$ (MHz) | $\lambda/2\pi$ (MHz) | $2(\lambda - \kappa/2)$ (ns$^{-1}$) | $\langle n \rangle$ at 20 ns | $t_{\mathrm{sat}}$ (ns) |
|---|---|---|---|---|---|
| 0 | 0.00 | 50.00 | 0.503 | $7.3\times10^{3}$ | 11.5 |
| 0.5 | 25.00 | 43.30 | 0.419 | $1.9\times10^{3}$ | 13.0 |
| 0.8 | 40.00 | 30.00 | 0.251 | $1.6\times10^{2}$ | 18.2 |
| 0.9 | 45.00 | 21.79 | 0.148 | $4.4\times10^{1}$ | 25.3 |
| 0.95 | 47.50 | 15.61 | 0.071 | $2.1\times10^{1}$ | 38.6 |
| 0.980 | 48.99 | 10.00 | 0 | $1.3\times10^{1}$ | 131.3 |

At the last row the closed-system exponent equals the amplitude loss rate $\kappa/2$. The deterministic net growth rate vanishes. Vacuum-noise refill then produces linear asymptotic occupation growth. In the first rows, the reported occupation at 20 ns is the linear-model extrapolation when that time exceeds the assumed saturation time.

**Data availability**

The code accompanying the bare-vacuum calculations is available at https://github.com/alexandermtishin-art/Intrinsic-Geometric-Energy-of-Driven-Quantum-Evolution. Figure 2 is generated directly from Eqs. (11a)–(12). The supplementary script two_fidelity_threshold.py verifies the inversion and generates Fig. 3.

**Declaration of generative AI use**

During the preparation of this work the author used generative AI tools to assist with language editing, notation consistency, reference cross-checking, and Python code development. The author reviewed and edited the output and takes full responsibility for the content of the publication.